\documentclass{cjaa}                   

\usepackage{graphicx}                   
\input{epsf.sty}                        
\input{psfig.sty}                       

\begin{document}

   \title{Observations of Rapid Disk-Jet Interaction in GRS~1915+105}

   \volnopage{{\bf preprint}, Vol.5 (2005) {\bf Suppl.}, 57-62}      
   \setcounter{page}{1}          

   \author{David M. Rothstein
      \inst{1}\mailto{}
   \and Stephen S. Eikenberry
      \inst{2}
   \and Keith Matthews
      \inst{3}
      }

   \institute{Department of Astronomy, Cornell University, 514
Space Sciences Building, Ithaca, NY, USA 14853\\
             \email{droth@astro.cornell.edu}
        \and
             Department of Astronomy, University of Florida, 211
Bryant Space Science Center, Gainesville, FL, USA 32611\\
        \and
             California Institute of
Technology, Downs Laboratory, MS 320-47, Pasadena, CA, USA 91125\\
          }

   \date{(Proceedings of the 5th Microquasar Workshop, Beijing, June 2004)}

   \abstract{
We present evidence that $\sim 30$ minute episodes of jet formation in GRS~1915+105 may sometimes be a superposition of smaller, faster phenomena.  Based on observations in 2002 July using the {\it Rossi X-ray Timing Explorer} and the Palomar
5 meter telescope, we show that GRS~1915+105 sometimes produces $\sim 30$ minute infrared flares that can entirely be explained as a combination of small ($\sim 150$ second) flares, one for each oscillation in the accompanying X-ray light curve.  We discuss the differences between these observations and similar ones in 1997 August and conclude that an X-ray ``trigger spike'' seen during each cycle in 1997 is a key ingredient for large ejections to occur on $\sim 30$ minute timescales in this source.
   }

   \authorrunning{Rothstein, Eikenberry \& Matthews}            
   \titlerunning{Rapid Disk-Jet Interaction in GRS~1915+105}  

   \maketitle

%
%
\section{Introduction}           
\label{sect:intro}

GRS~1915+105 is the most active microquasar in our Galaxy, with variability in many different wavebands and on many different timescales.  On the largest scales, resolved, bipolar relativistic ejections are occasionally observed in radio maps (e.g. Mirabel \& Rodr\'\i guez 1994) and evolve on timescales of weeks (the ``Class A'' ejections; Eikenberry et al. 2000).  However, GRS~1915+105 also exhibits smaller ``class B'' radio and infrared flares on $\sim 30$
minute timescales.  These events are also thought to correspond to jet ejection, and multiwavelength studies of their properties have proven particularly useful for understanding jet formation in this source (Pooley \& Fender 1997, Eikenberry et al. 1998, Mirabel et al. 1998).

In particular, observations in 1997 August by Eikenberry et al. (1998) showed a one-to-one correspondence between repeating X-ray variability cycles and infrared flares on timescales of $\sim 30$ minutes.  This is consistent with a picture in which emptying and refilling of the X-ray emitting inner accretion disk (e.g. Belloni et al. 1997a,b) triggers the ejection of material into a jet, which radiates through synchrotron emission to produce a flare (Fender et al. 1997; Pooley \& Fender 1997).  These observations were among the first to show the intimate link between accretion disk evolution and relativistic jet formation in any black hole system.

Here, we present multiwavelength observations of GRS~1915+105 during 2002 July that expand on this simple picture of ``class B'' events and shed new light on the disk-jet interaction (Rothstein, Eikenberry \& Matthews 2005).

   \begin{figure}
   \centering
   \includegraphics[width=10cm]{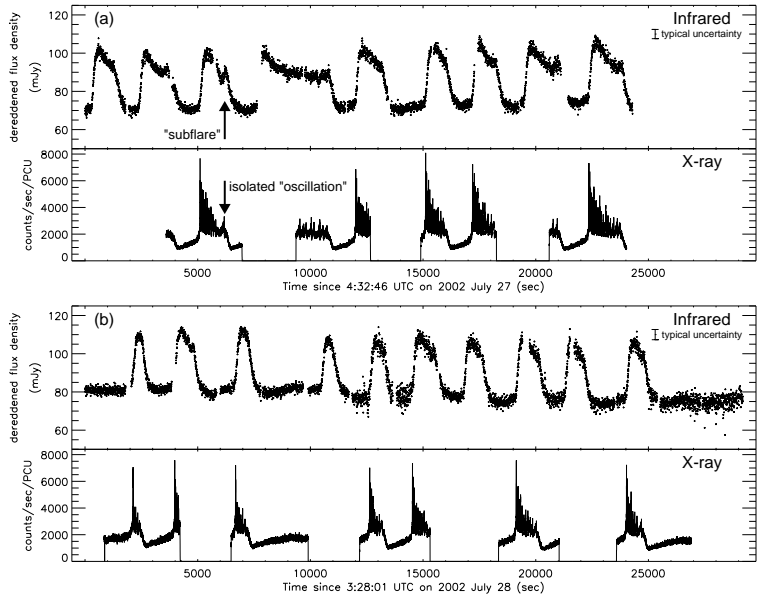}
\caption{\small Simultaneous infrared and X-ray light curves of GRS~1915+105
on (a) 2002 July 27 and (b) 2002 July 28.  Both light curves are at
1-second resolution, but the infrared data have been smoothed to
5-second resolution in this plot.  The infrared data were taken in the
K ($2.2 \mu$m) band and have been dereddened by 3.3 magnitudes;
typical 1-second $\pm$~$1 \sigma$ uncertainties are shown.  The X-ray data  are in the $\sim$ 2--100 keV energy range and are plotted in units of
counts per second per Proportional Counter Unit (PCU) of the RXTE PCA
instrument.  The ``subflare'' and isolated ``oscillation'' in (a) are discussed in the text.  This figure is taken from Rothstein, Eikenberry \& Matthews (2005).}
   \label{fig:lcurve}
   \end{figure}

\section{Observations}
\label{sect:Obs}

We obtained infrared observations of GRS~1915+105 in the K ($2.2 \mu$m) band, using the Palomar Observatory 5 meter Hale telescope.
X-ray observations (with coverage between $\sim$
2--100 keV) were obtained using the
Proportional Counter Array (PCA) on the {\it Rossi X-Ray Timing
Explorer} (RXTE).  Standard data reduction procedures were used for both the infrared and X-rays; details are presented in Rothstein, Eikenberry \& Matthews (2005).

Fig. \ref{fig:lcurve} shows infrared and X-ray light curves from July 27-28, the two nights for which we had multiwavelength coverage of the $\sim 30$ minute events.  As in the 1997 August observations of Eikenberry et al. (1998), there is clearly a one-to-one correspondence between infrared flares and periods of X-ray oscillation in the 2002 data.  However, there are also some key differences, discussed below.

\section{Infrared/X-ray Relationship}
\label{sect:infrared}

The infrared flares in 2002 are several times weaker than those in 1997 ($\sim 30$ mJy dereddened, as opposed to $\sim$ 100--200 mJy).  Furthermore, the duration of each flare in 2002 closely matches the duration of its accompanying X-ray oscillation period, while the flares in 1997 peaked and sometimes decayed back to their quiescent level before the X-rays stopped oscillating.

These differences can be explained if we assume that the 1997 infrared flares consist primarily of single, large ejections once per $\sim 30$ minute cycle, while the 2002 events consist primarily of blended-together faint flares associated with the individual X-ray oscillations, with no large ejection superimposed on them.

This interpretation is consistent with a faint infrared ``subflare'' observed on July 27 that is associated with an isolated oscillation in the X-ray light curve (Fig. \ref{fig:lcurve}a); if we assume that each X-ray oscillation has a similar or slightly stronger subflare associated with it, the subflares would blend together to produce roughly as much infrared emission as observed in each entire $\sim 30$ minute event.  We have tested this idea by attempting to simulate the infrared light curve directly, and we have found that a relatively simple model does a good job reproducing the observations.  Furthermore, our interpretation is consistent with a reanalysis of the 1997 observations by Eikenberry et al. (2000), who found that a faint $\sim 20$ mJy ``infrared excess'' seen during part of these observations could also be explained by a pileup of faint infrared flares associated with individual X-ray oscillations.  However, in 2002 it is not just an ``infrared excess,'' but rather the entire flare, which we can explain by this pileup process.

   \begin{figure}
   \centering
   \includegraphics[width=12cm]{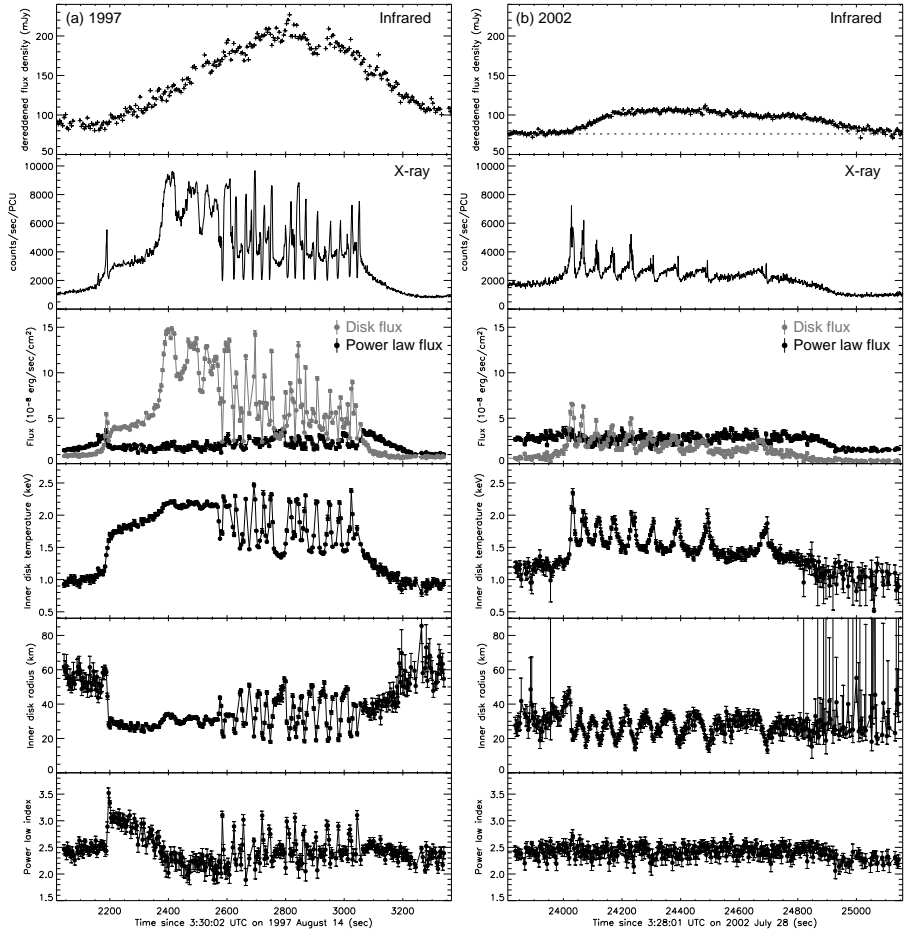}
\caption{\small Comparison of one cycle of jet formation in GRS~1915+105 from
observations in (a) 1997 and (b) 2002.  The top two panels in each
column show the infrared and X-ray light curves, respectively (at
1-second resolution, but the infrared data have been smoothed to
5-second resolution in this plot); the dotted horizontal line in (b)
shows the baseline infrared level.  The bottom four panels show the
results of our X-ray spectral fitting at 4-second resolution, plotted
along with $\pm$ $1 \sigma$ uncertainties; these include the
unabsorbed 2--25 keV flux from the multitemperature disk
blackbody and power law components of the spectrum, the temperature at
the inner edge of the accretion disk, the radius of the inner edge of
the accretion disk, and the power law index.  This figure is modified from Rothstein, Eikenberry \& Matthews (2005).}
   \label{fig:specfit}
   \end{figure}

\section{X-ray Spectral Evolution}
\label{sect:xray}

The X-ray evolution during the 2002 observations is similar to what was seen in 1997; both consist of $\sim 30$ minute cycles of ``dips'' followed by a period of oscillations.  The details, however, are different; in the classification of Belloni et al. (2000), GRS~1915+105 was in the ``class $\beta$'' state in 1997 and the ``class $\alpha$'' state in 2002.

In Fig. \ref{fig:specfit}, we show one cycle of jet formation from each of the 1997 and 2002 observations.  We fit the X-ray spectra during these observations using a standard model for black hole candidates, consisting of a multitemperature disk blackbody (e.g. Mitsuda et al. 1984) plus a power law, both modified by fixed hydrogen absorption.  Plots of several parameters from our fits (including the 2--25 keV disk and power law flux, the inner disk temperature, the inner disk radius, and the power law index) are shown in Fig. \ref{fig:specfit}, plotted at 4-second resolution.

  \begin{figure}
   \centering
   \includegraphics[width=8cm]{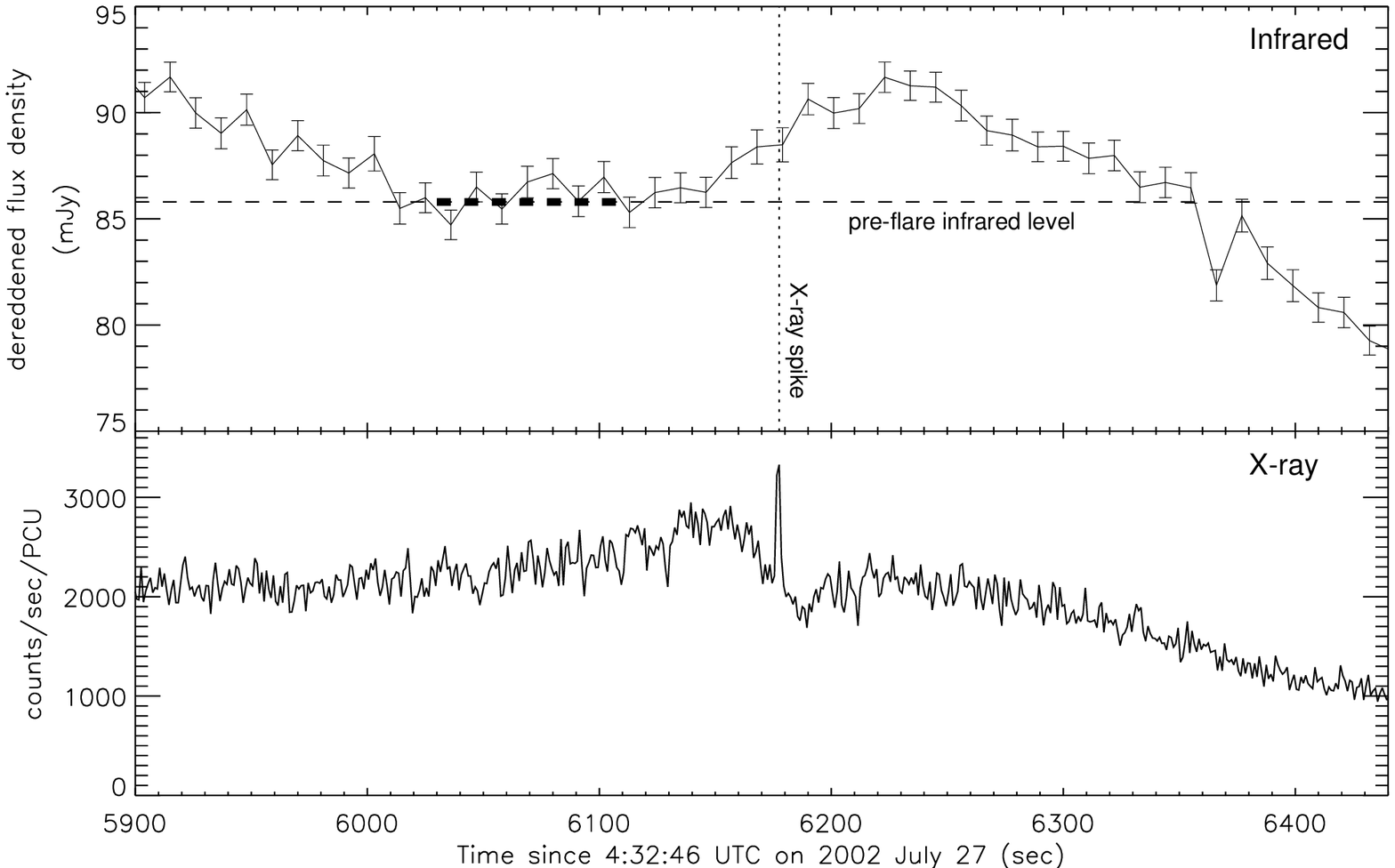}
\caption{\small Simultaneous infrared and X-ray light curves of GRS~1915+105
at the time of the infrared ``subflare'' observed on 2002 July 27
(Figure \ref{fig:lcurve}a).  The X-ray data are at 1-second
resolution, and the infrared data have been smoothed to 11-second
resolution to increase the signal-to-noise ratio; they are plotted
along with their $\pm$ $1 \sigma$ uncertainties.  The dotted vertical
line marks the time of the X-ray spike, and the dashed horizontal line
marks the median value of the infrared light curve calculated between
6,030 and 6,110 seconds (the bold region of the dashed line).  The
subflare appears to rise above this value \mbox{$\sim 30$} seconds
before the X-ray spike begins.  This figure is taken from Rothstein, Eikenberry \& Matthews (2005).}
   \label{fig:subflare}
   \end{figure}

The main difference between the 1997 and 2002 X-ray cycles is the appearance of a sharp ``trigger spike'' in the 1997 data (at $\sim 2200$ seconds in Fig. \ref{fig:specfit}a).  This spike, although only several seconds in duration, signals an abrupt change in the X-ray spectral properties of the source and leads into a period of a few hundred seconds where the inner accretion disk has ``turned on'' (as evidenced by the increase in inner disk temperature and disk flux and decrease in inner disk radius) but before the fast X-ray oscillations have begun.  In 2002, there is also a sharp spike which signals the beginning of the accretion disk activity, but it differs from the 1997 trigger spike in that there is no delay between it and the subsequent oscillations; in fact, its spectral properties are very similar to those of the oscillations, and it appears to simply be the first in a long series of similar events.  Because it is clear from Fig. \ref{fig:specfit}a that the large infrared flare in 1997 begins around the time of the trigger spike (and before the fast oscillations), we conclude that the trigger spike is associated with the large ejections that occur in 1997 but not in 2002.  This confirms previous suggestions (e.g. Eikenberry et al. 1998, Mirabel et al. 1998) that the trigger spike is an important signature of large ejections in GRS~1915+105.

Another interesting difference between the 1997 and 2002 X-ray behavior is in the evolution of the power law component; as can be seen from Fig. \ref{fig:specfit}, the power law index is much more variable in 1997 than in 2002, in particular during the fast X-ray oscillations.  We have examined individual X-ray spectra during this time period and found that the hard X-rays ($> 20$ keV) do in fact change their spectrum dramatically in 1997 while remaining relatively constant in 2002 --- however, we cannot rule out the possibility that the attribution of this change to the power law index is an artifact of our fitting procedure (since inaccuracies in the multitemperature disk blackbody model which dominates the soft X-ray spectrum could conceivably lead to spurious changes in the fitted values of the power law parameters).

\section{What Causes the Infrared Subflares?}
\label{sect:subflares}

We show in Fig. \ref{fig:subflare} a closer view of the isolated ``subflare'' from the first night of our observations.  The subflare is associated with a spike in the X-ray light curve, but this spike is very different from the trigger spike in 1997; it is not accompanied by any abrupt changes in the X-ray spectrum indicating that the state of the system has changed.  The X-ray behavior during this event, including the X-ray spike, is similar to that seen in most of the other X-ray oscillations in 2002.

The subflare clearly peaks after the X-ray spike with which it is associated, but its initial rise does seem to begin $\sim 30$ seconds {\it before} the spike.  A more extreme version of this behavior was seen in several isolated pairs of X-ray oscillations and faint infrared flares observed by Eikenberry et al. (2000) in 1998 July.  An ``outside-in'' origin for these events was suggested by Eikenberry et al. (2000), in which the flare occurs somewhere outside of the inner accretion disk (either in the outer part of the disk or as an internal shock within the jet) and then propagates inward to cause a disturbance in the X-rays emitted in the inner region.

We cannot directly rule out a thermal origin for the 2002 subflare, but the events seen by Eikenberry et al. (2000) were thought to be synchrotron in origin, and that is the most likely cause for these events as well.  In addition, 15 GHz observations at the Ryle Telescope several hours before the second night of our run show $\sim$ 10--20 mJy radio flares (G. Pooley 2002, private communication), suggesting that the $\sim 30$ minute events we saw had a nonthermal component and that each small subflare is nonthermal as well.  However, we cannot completely rule out a scenario in which the radio flares are single events produced once per $\sim 30$ minute cycle and the repeating, superimposed infrared subflares have little associated radio emission.  In fact, observations by Ueda et al. (2002) during the ``class $\alpha$'' state show at least one radio flare that, unlike the infrared flares we saw, decays before the end of its accompanying X-ray oscillation period; therefore, it is unlikely to be associated with the oscillations.  Further simultaneous observations at X-ray, radio and infrared wavelengths are needed to resolve this issue.

\section{Discussion}
\label{sect:discussion}

The observations presented here show the importance of the X-ray ``trigger spike'' for jet formation on $\sim 30$ minute timescales in GRS~1915+105.  With it, large infrared flares are produced; without it, they are not.  More broadly, these observations show the intimate connection between accretion disk evolution and jet formation in this source; it appears that a faint infrared flare is produced every time an X-ray oscillation occurs, and that these flares can blend together to produce a significant amount of infrared emission on $\sim 30$ minute timescales that might otherwise be mistaken for a single, large ejection event.  This picture is in contrast to the one presented by Klein-Wolt et al. (2002), who have argued that the radio flares in GRS~1915+105 are produced via a continuous ejection of material during the long, spectrally hard ``dips'' in the X-ray light curve, and not during the periods of accretion disk activity.  In the infrared, however, it is clear from our 2002 observations that the observed flares must be directly related to the oscillations in the accretion disk, and the differences between our 1997 and 2002 observations strongly suggest that the trigger spike, not the dip, is the main part of the X-ray light curve which controls the large flaring behavior.

\begin{acknowledgements}
We thank the staff at Palomar Observatory and the members of the {\it
Rossi X-Ray Timing Explorer} team for their help with these
observations.  We also thank G. Pooley for providing the radio data
from the Ryle Telescope, and R. Lovelace and M. Tagger for useful
discussions.  D.~M.~R. is supported by a National Science Foundation
Graduate Research Fellowship.  S.~S.~E. is supported in part by an NSF
CAREER award (NSF-9983830).
\end{acknowledgements}

\label{lastpage}


\begin{thebibliography}{99}

\bibitem[1997a]{Belloni97a} Belloni T., M\'{e}ndez M., King A.~R., van der Klis M., van Paradijs J. 1997a, \apj, 479, L145

\bibitem[1997b]{Belloni97b} Belloni T., M\'{e}ndez M., King A.~R., van der Klis M., van Paradijs J. 1997b, \apj, 488, L109

\bibitem[2000]{Belloni2000} Belloni T., Klein-Wolt M., M\'{e}ndez M., van der Klis M., van Paradijs J. 2000, \aap, 355, 271

\bibitem[1998]{Eiken98} Eikenberry S.~S., Matthews K., Morgan E.~H., Remillard R.~A., Nelson R.~W. 1998, \apj, 494, L61

\bibitem[2000]{Eiken2000} Eikenberry S.~S., Matthews K., Muno M., Blanco P.~R., Morgan E.~H., Remillard R.~A. 2000, \apj, 532, L33

\bibitem[1997]{Fender97} Fender R.~P., Pooley G.~G., Brocksopp C., Newell S.~J. 1997, \mnras, 290, L65

\bibitem[2002]{KleinWolt2002} Klein-Wolt M., Fender R.~P., Pooley G.~G., Belloni T., Migliari S., Morgan E.~H., van der Klis M. 2002, \mnras, 331, 745

\bibitem[1994]{Mirabel94} Mirabel I.~F., Rodr\'\i guez~L. F. 1994, Nature, 371, 46

\bibitem[1998]{Mirabel98} Mirabel I.~F., Dhawan V., Chaty S., Rodr\'\i guez L.~F., Mart\'\i~J., Robinson C.~R., Swank J., Geballe T.~R. 1998, \aap, 330, L9

\bibitem[1984]{Mitsuda} Mitsuda K., et al. 1984, \pasj, 36, 741

\bibitem[1997]{PooleyFender97} Pooley G.~G., Fender R.~P. 1997, \mnras, 292, 925

\bibitem[2005]{rothsteinapj} Rothstein D.~M., Eikenberry S.~S., Matthews K. 2005, \apj, 626, 991

\bibitem[2002]{Ueda} Ueda Y., et al. 2002, \apj, 571, 918

\end{thebibliography}
\end{document}